\documentclass[aps,preprint]{revtex4}%
\usepackage{amsfonts}
\usepackage{amsmath}
\usepackage{amssymb}
\usepackage{subfigure}
\usepackage{graphicx}%

\begin{document}
\title{Thermodynamics and overcharging problem in the extended phase spaces of charged AdS
black holes with cloud of strings and quintessence under charged particle absorption}
\author{Jing Liang$^{a,b}$}
\email{jingliang@stu.scu.edu.cn}

\author{Benrong Mu$^{a,b}$}
\email{benrongmu@cdutcm.edu.cn}

\author{Jun Tao$^{b}$}
\email{taojun@scu.edu.cn}

\affiliation{$^{a}$ Physics Teaching and Research section, College of Medical Technology,
Chengdu University of Traditional Chinese Medicine, Chengdu, 611137,
PR China}
\affiliation{$^{b}$Center for Theoretical Physics, College of Physics, Sichuan University, Chengdu, 610064, PR China}

\begin{abstract}
The thermodynamics and overcharging problem in the RN-AdS black hole with the cloud of strings and quintessence are investigated by absorption of scalar particle and fermion in the extend phase space. The cosmological constant is treated as the pressure with a conjugate volume. Besides, the parameters related to quintessence and cloud of strings are treated as thermodynamic variables. Finally we find the first law of thermodynamics is satisfied and the second law of thermodynamics is indefinite. Furthermore, the near-extremal and extremal black holes can not be overcharged.

\end{abstract}

\maketitle
\tableofcontents{}

\bigskip{}

%\affiliation{Center for Theoretical Physics, College of Physical Science and Technology,
%Sichuan University, Chengdu, 610064, PR China}

%\affiliation{Center for Theoretical Physics, College of Physical Science and Technology,
%Sichuan University, Chengdu, 610064, PR China}

\section{Introduction}
Since Stephen Hawking proved that black holes have quantum radiation with a temperature $T=\frac{\kappa}{2\pi}$ \cite{intro-Hawking:1974sw}, it is believed that a black hole can be treated as a thermodynamic system. Until now, there have been many work on the thermodynamics of black holes, such as the four laws of thermodynamics \cite{intro-Bekenstein:1972tm,intro-Bekenstein:1974ax}, phase transitions \cite{intro-Zeng:2016aly} and quantum effects \cite{intro-Zeng:2008zzc,intro-Zeng:2015nfa}. In general, the first law of thermodynamics of black hole is written as
\begin{equation}
dM=TdS+\varPhi dQ.
\label{eqn:dM1}
\end{equation}
Where $M$ is the mass, $T$ is the Hawking temperature of the black hole, $S$ is the entropy, $\varPhi$ is the electric potential and $Q$ is the electric charge. The mass is usually interpreted as the enthalpy \cite{intro-Kastor:2009wy}. It is clear that there is no $PdV$ term in Eq. $\left(\ref{eqn:dM1}\right)$, which corresponds to the change in volume under pressure $P$. When the cosmological constant, $\varLambda$, is treated as the pressure of
the black hole \cite{intro-Dolan:2011xt,intro-Kubiznak:2012wp,intro-Cvetic:2010jb,intro-Caceres:2015vsa,intro-Hendi:2012um,intro-Pedraza:2018eey} and the volume of the black hole is defined as the thermodynamic variable conjugate to the pressure \cite{intro-Dolan:2010ha}, Eq. $\left(\ref{eqn:dM1}\right)$ is modified as
\begin{equation}
dM=TdS+VdP+\varPhi dQ.
\label{eqn:dM2}
\end{equation}
The relations between $P$, $\varLambda$ and $V$ are $P=-\frac{\varLambda}{8\pi}$, $V=\left(\frac{\partial M}{\partial P}\right)_{S,Q}$. Recently the existence of a gravitationally repulsive interaction at a global scale (cosmic dark energy) was confirmed by high-precision observations \cite{intro-Ade:2013sjv}. It is founded that one type of dark energy models produces some gravitational effect when it surrounds black holes. For this type of dark energy, the equation of state parameters is in the interval $[-1,-\frac{1}{3}]$ \cite{intro-Saleh:2011zz}. This type of dark energy models is called quintessence dark energy or quintessence for short. In this case, the first law of thermodynamics is given by \cite{intro-Li:2014ixn}
\begin{equation}
dM=TdS+VdP+\varPhi dQ-\frac{1}{2r_{+}^{3\omega_{q}}}d\alpha,
\label{eqn:dM3}
\end{equation}
where $\alpha$ is a positive normalization factor. There has been much interest in studying the physics of black holes surrounded by quintessence \cite{intro-Singh:2020tkf,intro-Haldar:2020jmt, intro-Chen:2020rov,intro-Hong:2019yiz,intro-Moinuddin:2019mzf,intro-Toledo:2019mlz,intro-Liu:2017baz,intro-Ghaderi:2016dpi,intro-Fernando:2014rsa,intro-Fernando:2014wma,intro-Guo:2019hxa,intro-Nandan:2016ksb,intro-Malakolkalami:2015cza,intro-WangChun-Yan:2012tcg,intro-Xi:2008ce,intro-Harada:2006dv, intro-Kiselev:2002dx}.

According to string theory, nature can be represented by a set of extended objects (such as one-dimensional strings) rather than point particles. Therefore, it is of great importance to understand the gravitational effects caused by a set of strings, which can be achieved by solving Einstein's equations with a finite number of strings. The results obtained by the Letelier show that the existence of cloud of strings will produce a global origin effect that related to a solid deficit angle. The solid deficit angle depends on the parameters that determine the existence of the cloud \cite{intro-Letelier:1979ej}. Therefore, the existence of cloud of strings will have an impact on black holes. In this case, the first law of thermodynamics takes on the form as
\begin{equation}
dM=TdS+VdP+\varPhi dQ-\frac{r_{+}}{2}da,
\label{eqn:dM4}
\end{equation}
where $a$ is the state parameter of cloud of strings. There have been many interesting research results about the black holes surrounded by the cloud of strings \cite{intro-Li:2020zxi,intro-Cai:2019nlo,intro-Toledo:2019szg,intro-Ghaffarnejad:2018tpr,intro-Graca:2016cbd,intro-Ghosh:2014dqa}. As noted in \cite{intro-Toledo:2019amt}, considered that the parameters related to the cloud of string and quintessence are extensive thermodynamic parameters. Then the first law of thermodynamics of black hole is modified as
\begin{equation}
dM=TdS+VdP+\varPhi dQ-\frac{1}{2r_{+}^{3}\omega_{q}}d\alpha-\frac{r_{+}}{2}da.
\label{eqn:dM5}
\end{equation}
There has been much interest in deducing and discussing the physical properties of various black holes when they are surrounded by cloud of strings and quintessence \cite{intro-Chabab:2020ejk,intro-Sakti:2019iku,intro-Ma:2019pya, intro-Toledo:2019amt,intro-Toledo:2018pfy,intro-Toledo:2020xnt,intro-Toledo:2018hav}.

An important feature of a black hole is its horizon, no matter what matter can escape through it. There is a gravitational singularity in the center of the black hole, which hides in the event horizon. At the singularity, all the laws of physics break down. In order to avoid this phenomenon, Penrose proposed the weak cosmic censorship conjecture (WCCC) in 1969 \cite{intro-Penrose:1964wq,intro-Penrose:1969pc}. The WCCC claims that the singularity is always hidden in the event horizon and cannot be seen by the observer at the infinite distance. Although the WCCC's correctness is widely accepted, there is no complete evidence to prove it and people can only test its validity. The Gedanken experiment is an effective method to test the validity of the WCCC \cite{intro-Sorce:1974dst}. In this experiment, a test particle with sufficient energy, charge and angular momentum is thrown into the black hole. After the black hole absorbed the test particle, if the horizon is destroyed, the singularity becomes a naked singularity. In this case, the black hole is overcharged and the WCCC is violated. On the contrary, if the horizon isn't destroyed, the singularity is surrounded by the horizon. Consequently the black hole is not overcharged and the WCCC is valid. After this experiment was proposed, the validity of WCCC has been tested in various black holes by this experiment \cite{intro-Ying:2020bch,intro-Yang:2020czk,intro-Mu:2019bim,intro-Hu:2019lcy,intro-He:2019fti,intro-Liu:2020cji,intro-Hu:2020lkg,intro-Wang:2020osg,intro-Wang:2019jzz,intro-Chen:2020zps,intro-Hu:2020ccj,intro-Zeng:2019hux,intro-Zeng:2019aao,intro-Han:2019lfs,intro-Han:2019kjr,intro-Zeng:2019jrh,intro-Matsas:2007bj,intro-Hod:2008zza,intro-Isoyama:2011ea,intro-Richartz:2011vf,intro-Gim:2018axz}. On the other hand, the validity of the WCCC can be investigated through the Gedanken experiment by using test fields instead of test particles. The experiment was first proposed by Semiz \cite{intro-Semiz:2005gs}. This experiment is also used to study the validity of the WCCC in different space-times \cite{intro-Hong:2020zcf,intro-Yang:2020iat,intro-Bai:2020ieh,intro-Chen:2019nsr,intro-Chen:2018yah,intro-Gwak:2018akg,intro-Toth:2011ab,intro-Goncalves:2020ccm,intro-Jiang:2020btc,intro-Duztas:2018adf,intro-Duztas:2019mxr}.

In this paper, we investigate the thermodynamics and overcharging problem in the RN-AdS black hole with cloud of strings and quintessence by the charged particle absorption in the extend phase space. Due to the existence of the cloud of strings and quintessence, the constants relate to them are also taken into account in the calculation. The organizational structure of this paper is as follows. In Section \ref{sec:M}, we review the thermodynamics of the RN-AdS black hole with cloud of strings and quintessence. In Section \ref{sec:particle}, the absorptions of the scalar particle and fermion are discussed. In Section \ref{sec:T}, the first and the second laws of thermodynamics are investigated in the extended phase space. In Section \ref{sec:wccc}, the overcharging problem is tested in the near-extremal and extremal black holes. Our results are summarized in Section \ref{sec:con}.
\section{Black hole solution}
\label{sec:M}
The metric of a RN-AdS black hole surrounded by cloud of strings and quintessence in 4-dimensional space-time is given by \cite{intro-Singh:2020tkf,intro-Ma:2019pya,intro-Chabab:2020ejk}
\begin{equation}
ds^{2}=f(r)dt^{2}-\frac{1}{f(r)}dr^{2}-r^{2}\left(d\theta^{2}+sin^{2}\theta d\phi^{2}\right),
\label{eqn:ds2}
\end{equation}
with
\begin{equation}
f(r)=1-a-\frac{2M}{r}+\frac{Q^{2}}{r^{2}}-\frac{\alpha}{r^{3\omega_{q}+1}}-\frac{\text{\ensuremath{\Lambda}}r^{2}}{3}.
\end{equation}
In the above equation, $\Lambda$ is the cosmological constant related to the AdS space radius $l$ by $\Lambda=-3/l^{2}$, $M$ and $Q$ are constants respectively equal to the mass and charge of the black hole, respectively. $a$ is the integral constant caused by the cloud of strings and $\alpha$ is normalization constants related to the quintessence, with density $\rho_{q}$ as
\begin{equation}
\rho_{q}=-\frac{\alpha}{2}\frac{3\omega_{q}}{r^{3(\omega_{q}+1)}}.
\end{equation}
\begin{figure}[htb]
\begin{center}
\subfigure[{$\alpha=0.01$, $\omega_{q}=-2/3$.}]{
\includegraphics[width=0.45\textwidth]{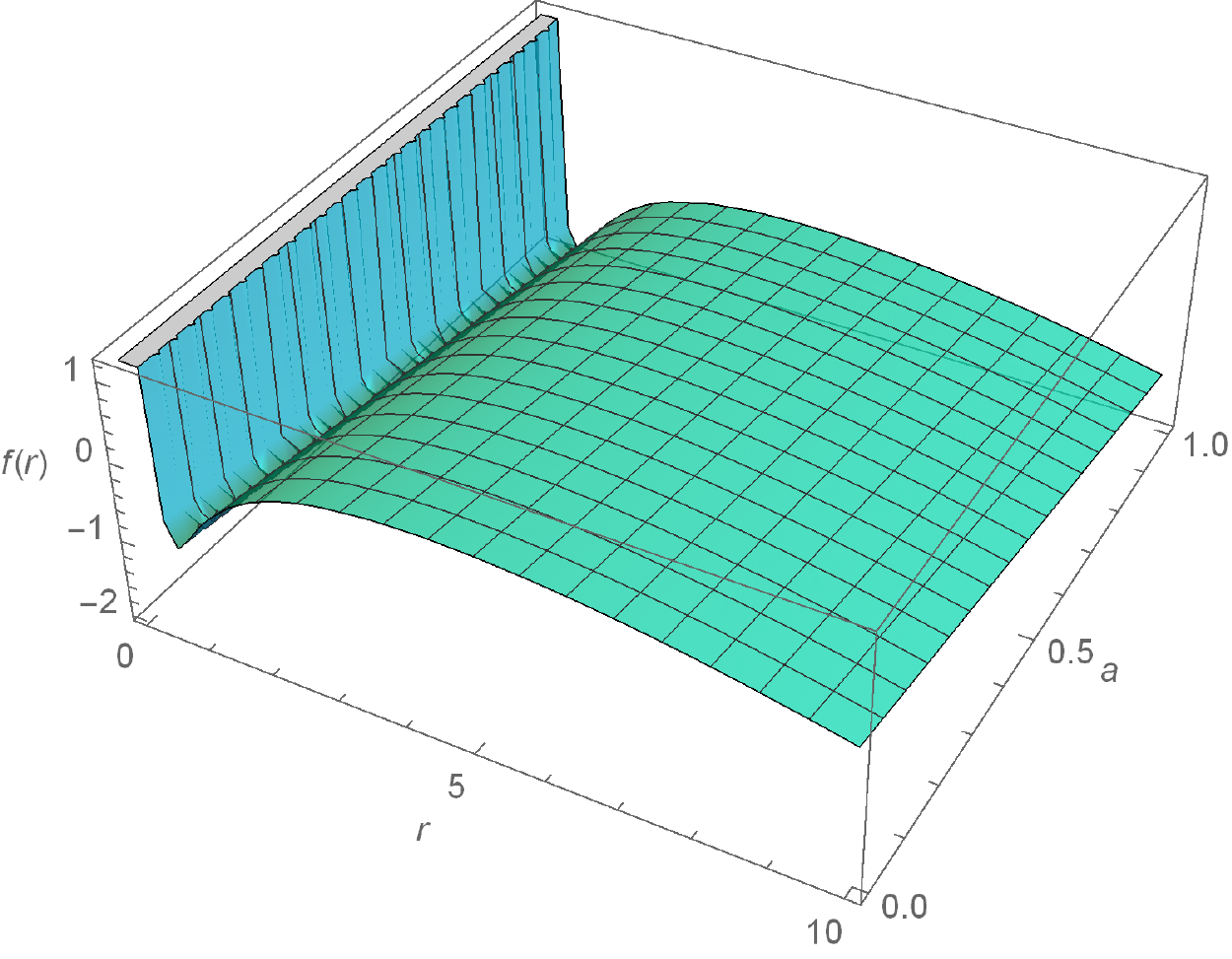}\label{fig:f1}}
\subfigure[{$\alpha=0.1$, $\omega_{q}=-1/2$.}]{
\includegraphics[width=0.45\textwidth]{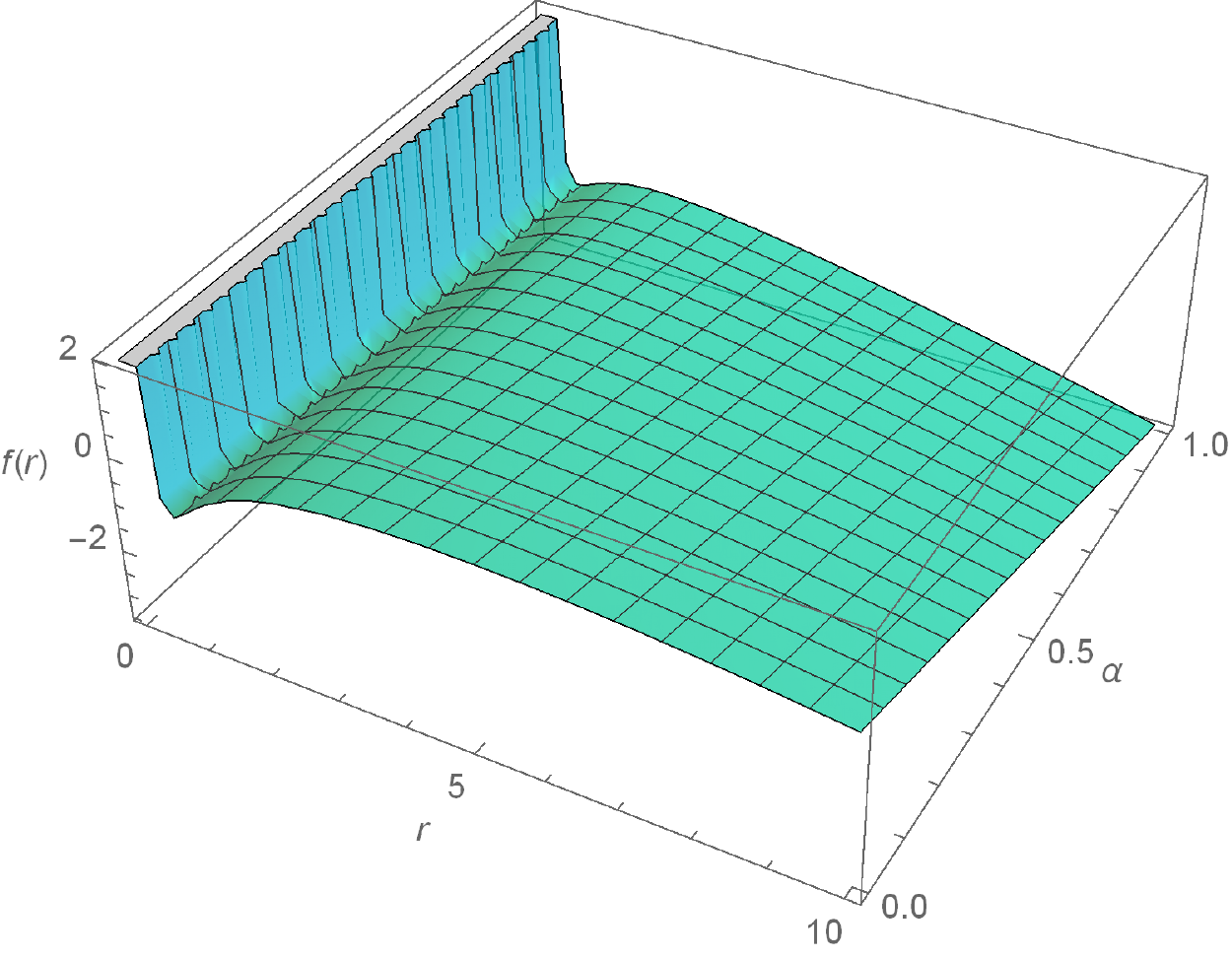}\label{fig:f2}}
\end{center}
\caption{The function $f(r)$ for different values of $a$, $\alpha$ and $\omega_{q}$. We choose $M=1$ and $Q=0.8$.}%
\label{fig:f}
\end{figure}
In Fig. \ref{fig:f}, the graphs of the function $f(r)$ are shown for different
values of the parameters $a$, $\alpha$ and $\omega_{q}$, which
represent the presence of the cloud of strings and the quintessence.

When it is the non-extremal black hole, the equation $f(r)=0$ has two positive real roots $r_+$ and $r_-$. The $r_+$ represents the radius of the event horizon. When it is the extremal black hole, $f(r)=0$ has only one root $r_+$. The mass of the black hole is
\begin{equation}
M=\frac{1}{2}(r-ar+\frac{Q^{2}}{r^{2}}-\frac{\alpha}{r^{3\omega_{q}}}+\frac{r^{3}}{l^{2}}).
\end{equation}
The Hawking temperature takes on the form as
\begin{equation}
T=\frac{f^{\prime}(r_{+})}{4\pi}=\frac{1}{4\pi}(\frac{2M}{r_{+}^{2}}-\frac{2Q^{2}}{r_{+}^{3}}+\frac{(3\omega_{q}+1)\alpha}{r_{+}^{3\omega_{q}+2}}+\frac{2r_{+}}{l^{2}}).
\label{eqn:T}
\end{equation}
Moreover, the entropy and the potential of the black hole are
\begin{equation}
S=\pi r_{+}^{2},
\label{eqn:entropy}
\end{equation}
\begin{equation}
\Phi=-A_{t}\left(r_{+}\right)=\frac{Q}{r_{+}}.
\label{eqn:potential}
\end{equation}
In previous studies, the cosmological constant is treated as a constant. Recently, the thermodynamic pressure of the black hole is introduced into the laws of thermodynamics. The cosmological constant is treated as a variable
related to pressure. The relationship between the cosmological constant and pressure is \cite{intro-Dolan:2011xt,intro-Kubiznak:2012wp,intro-Cvetic:2010jb,intro-Caceres:2015vsa,intro-Hendi:2012um,intro-Pedraza:2018eey}
\begin{equation}
P=-\frac{\Lambda}{8\pi}=\frac{3}{8\text{\ensuremath{\pi}}l^{2}}.
\label{eqn:P}
\end{equation}
The first law of thermodynamics in the extended phase space is written as
\begin{equation}
dM=TdS+VdP+\text{\ensuremath{\varphi}}dQ+\gamma\text{\ensuremath{d\alpha}}+\varkappa da,
\end{equation}
where
\begin{equation}
\gamma=-\frac{1}{2r_{+}^{3\omega_{q}}},\varkappa=-\frac{r_{+}}{2}.
\end{equation}
In the above equation, the volume is given by
\begin{equation}
V=\left(\frac{\partial M}{\partial P}\right)_{S,Q}=\frac{4\text{\ensuremath{\pi}}r_{+}^{3}}{3}.
\label{eqn:V}
\end{equation}
The mass of the black hole $M$ is defined as its enthalpy. Hence, the relationship between enthalpy, internal energy and pressure is \cite{intro-Cvetic:2010jb,intro-Kastor:2009wy}
\begin{equation}
M=U+PV.
\label{eqn:M}
\end{equation}

\section{Particles¡¯ absorption}
\label{sec:particle}
\subsection{Scalar particle¡¯s absorption}
\label{subsec:scalar}
In curved space-time, the motion of the charged scalar particle satisfies the Klein-Gordon equation
\begin{equation}
\frac{1}{\sqrt{-g}}(\frac{\partial}{\partial x^{\mu}}-\frac{iq}{\hbar}A_{\mu})[\sqrt{-g}g^{\mu\nu}(\frac{\partial}{\partial x^{\nu}}-\frac{iq}{\hbar}A_{\nu})]\Psi_{S}-\frac{m^{2}}{\hbar^{2}}\Psi_{S}=0,
\label{eqn:KG}
\end{equation}
where $\Psi_{S}$ is the scalar field, $m$ and $q$ are the mass and charge of the particle, respectively. Using the WKB approximation \cite{sca-Gillani:2011dj,sca-Ejaz:2013fla,sca-Sakalli:2015raa}, the wave function is written as
\begin{equation}
\Psi_{S}=exp(\frac{i}{\hbar}I+I_{1}+\mathcal{O}(\hbar)).
\label{eqn:Psi}
\end{equation}
Inserting Eq. $\left(\ref{eqn:Psi}\right)$ and the contravariant metric
components of the 4-dimensional black hole into the Klein-Gordon equation, we obtain
\begin{equation}
f^{-1}(\partial_{t}I-qA_{t})^{2}-f(\partial_{r}I)^{2}-\frac{1}{r^{2}}(\partial_{\theta}I)^{2}-\frac{1}{r^{2}sin^{2}\theta}(\partial_{\varphi}I)^{2}+m^{2}=0.
\label{eqn:K G}
\end{equation}
Considering the symmetry of space-time, it is necessary to carry out the separation of variables in the action
\begin{equation}
I=-\omega t+W(r)+S(\theta,\varphi),
%\label{eqn:}
\end{equation}
where $\omega$ is the energy of the absorbed scalar particle. Substituting the above separated action into Eq. $\left(\ref{eqn:K G}\right)$ and simplifying it yields
\begin{equation}
\partial_{r}W=\pm\frac{\sqrt{(\omega-\frac{4qQ}{r})^{2}+[m^{2}-\frac{1}{r^{2}}(\partial_{\theta}S)^{2}-\frac{1}{r^{2}sin^{2}\theta}(\partial_{\varphi}S)^{2}]f}}{f}.
%\label{eqn:}
\end{equation}
In the above equation, $+(-)$ indicates the situation of ingoing(outgoing) particles. The negative sign is ignored when we suppose the particles are completely absorbed by the black hole \cite{sca-Gwak:2017kkt}. Defining $p^{r}=\partial_{r}I=\partial_{r}W$ and the above equation is modified as
\begin{equation}
p^{r}=g^{rr}p_{r}=\sqrt{(\omega-\frac{4qQ}{r})^{2}+[m^{2}-\frac{1}{r^{2}}(\partial_{\theta}S)^{2}-\frac{1}{r^{2}sin^{2}\theta}(\partial_{\varphi}S)^{2}]f}.
\label{eqn:gp}
\end{equation}
Here we consider the situation of absorbed particles near the horizon, which means $f(r)\rightarrow0$. Then Eq. $\left(\ref{eqn:gp}\right)$ is simplified to
\begin{equation}
p^{r}=\omega-q\Phi,
\label{eqn:pr1}
\end{equation}
where $\Phi=\frac{Q}{r_{+}}$ is the electric potential. The Eq. $\left(\ref{eqn:pr1}\right)$ is the relationship between the momentum, the energy and the charge of the ingoing particle. When $\omega < q\Phi$, the energy of the black hole flows out the event horizon and the superradiation happens. When $\omega=q\Phi$, the energy of the black hole does not change. In the discussion of this paper, it is assumed that $\omega \geq q\Phi$, which implies superradiation does not occur. The Eq. $\left(\ref{eqn:pr1}\right)$ plays an important role in the discussing of the thermodynamics of black holes and is recovered by fermion's absorption in the next section.

\subsection{Fermion¡¯s absorption}
\label{subsec:fer}
In curved space-time, the motion of the charged fermion particle obeys the Dirac equation
\begin{equation}
i\gamma^{\mu}(\partial_{\mu}+\Omega_{\mu}-\frac{i}{\hbar}qA_{\mu})\Psi_{F}+\frac{m_0}{\hbar}\Psi_{F}=0,
\label{eqn:Dirac}
\end{equation}
where $m$ and $q$ are the mass and charge of the fermion, respectively. $\Omega_{\mu}\equiv\frac{i}{2}\omega_{\mu}{}^{ab}\varSigma_{ab}$, $\omega_{\mu}^{ab}$ is the spin connection defined by the normal connection and the tetragonal ${\displaystyle e^{\lambda}}_{b}$. The relation between the spin connection, the normal connection and the tetragonal is
\begin{equation}
{\displaystyle {\displaystyle \omega_{\mu}}^{a}}_{b}=e_{\nu}{\displaystyle ^{a}}e^{\lambda}{\displaystyle _{b}}\Gamma_{\mu\lambda}^{\nu}-{\displaystyle e^{\lambda}}_{b}\partial_{\mu}e_{\lambda}{\displaystyle ^{a}}.
%\label{eqn:}
\end{equation}
The Greek index rises and falls with the curved metric $g_{\mu\nu}$. The Latin index is dominated by flat metric $\eta_{ab}$. To construct the tetrad, it is necessary to use the following definition
\begin{equation}
g_{\mu\nu}=e_{\mu}{\displaystyle ^{a}}e_{\nu}{\displaystyle ^{b}}\eta_{ab},   \eta_{ab}=g_{\mu\nu}{\displaystyle e^{\mu}}_{a}{\displaystyle e^{\nu}}_{b},
{\displaystyle e^{\mu}}_{a}e_{\nu}{\displaystyle ^{a}}=\delta_{\nu}^{\mu},
{\displaystyle e^{\mu}}_{a}e_{\mu}{\displaystyle ^{b}}=\delta_{a}^{b}.
%\label{eqn:}
\end{equation}
The Lorentz spinor generators are defined by
\begin{equation}
\varSigma_{ab}=\frac{i}{4}[\gamma^{a},\gamma^{b}],
\{\gamma^{\mu},\gamma^{\nu}\}=2\eta^{ab}.
%\label{eqn:}
\end{equation}
Then the $\gamma^{\mu}$ is constructed in curved space-time as
\begin{equation}
\gamma^{\mu}={\displaystyle e^{\mu}}_{a}\gamma^{a},
\{\gamma^{\mu},\gamma^{\nu}\}=2g^{\mu\nu}.
%\label{eqn:}
\end{equation}
For a fermion with a spin of $1/2$, its wave function must be described as both spin-up and spin-down. We first describe the spin-up wave function. The wave function takes on the form as
\begin{equation}
\Psi_{F\uparrow}=\text{\ensuremath{\left(\begin{array}{c}
A\\
0\\
B\\
0
\end{array}\right)}}exp\left(\frac{i}{\hbar}I_{\uparrow}(t,r,\theta,\varphi)\right).
%\label{eqn:}
\end{equation}
where $A$, $B$ and $I$ are the functions of $t$, $r$, $\theta$, $\phi$. For the metric $\left(\ref{eqn:ds2}\right)$, we chose
\begin{equation}
e_{\mu}{}^{a}=diag\left(\sqrt{f},1/\sqrt{f},r,rsin\theta\right).
%\label{eqn:}
\end{equation}
Then the $\gamma^{\mu}$ matrices is written as
\begin{equation}
\begin{aligned}
&\gamma^{t}=\frac{1}{\sqrt{f\left(r\right)}}\left(\begin{array}{cc}
i & 0\\
0 & -i
\end{array}\right),\gamma^{\theta}=r\left(\begin{array}{cc}
0 & \sigma^{1}\\
\sigma^{1} & 0
\end{array}\right),\\
&\gamma^{r}=\sqrt{f\left(r\right)}\left(\begin{array}{cc}
0 & \sigma^{3}\\
\sigma^{3} & 0
\end{array}\right),\gamma^{\varphi}=rsin\theta\left(\begin{array}{cc}
0 & \sigma^{2}\\
\sigma^{2} & 0
\end{array}\right),\\
\end{aligned}
%\label{eqn:}
\end{equation}
where $\sigma^{i}$ is the Pauli matrices which given by
\begin{equation}
\sigma^{1}=\left(\begin{array}{cc}
0 & 1\\
1 & 0
\end{array}\right),\sigma^{2}=\left(\begin{array}{cc}
0 & -i\\
i & 0
\end{array}\right),\sigma^{3}=\left(\begin{array}{cc}
1 & 0\\
0 & -1
\end{array}\right).
%\label{eqn:}
\end{equation}
Inserting the spin connection, wave function and gamma matrices into the Dirac equation, we obtain
\begin{equation}
-iA\frac{1}{\sqrt{f}}\left(\partial_{t}I_{\uparrow}-qA_{t}\right)-B\sqrt{g}\partial_{r}I_{\uparrow}+Am_{0}=0,
\label{eqn:A}
\end{equation}
\begin{equation}
-iB\frac{1}{\sqrt{f}}\left(\partial_{t}I_{\uparrow}-qA_{t}\right)-A\sqrt{g}\partial_{r}I_{\uparrow}+Bm_{0}=0,
\label{eqn:B}
\end{equation}
\begin{equation}
A\left[r\partial_{\theta}I_{\uparrow}+irsin\theta\partial_{\varphi}I_{\uparrow}\right]=0,
\label{eqn:A36}
\end{equation}
\begin{equation}
B\left[r\partial_{\theta}I_{\uparrow}+irsin\theta\partial_{\varphi}I_{\uparrow}\right]=0.
\label{eqn:B37}
\end{equation}
Eqs. $\left(\ref{eqn:A36}\right)$ and $\left(\ref{eqn:B37}\right)$ are simplified to one equation and yield $r^{2}\left(\partial_{\theta}I_{\uparrow}\right)^{2}+r^{2}sin^{2}\theta\left(\partial_{\varphi}I_{\uparrow}\right)^{2}=0$. In previous studies, the contribution of the angle part will not affect the results of tunneling radiation when the quantum gravity effects are not considered \cite{fer-Chen:2009bja}. Since the radial action is determined by the first two of the above four equations, we mainly focus on them. To solve the question we addressing, it is necessary to use the separation of variables
\begin{equation}
I_{\uparrow}=-\omega t+W\left(r\right)+\varTheta\left(\theta,\varphi\right).
\label{eqn:I}
\end{equation}
In the above equation, $\omega$ is the energy of the ingoing fermion. Inserting Eq. $\left(\ref{eqn:I}\right)$ into Eqs. $\left(\ref{eqn:A}\right)$ and $\left(\ref{eqn:B}\right)$, we obtain
\begin{equation}
f^{2}\left(\partial_{r}W\right)^{2}-\left(\omega-\frac{4qQ}{r}\right)^{2}-m_{0}^{2}f=0.
\label{eqn:fW}
\end{equation}
Simplify Eq. $\left(\ref{eqn:fW}\right)$, we have
\begin{equation}
\partial_{r}W=\pm\frac{\sqrt{\left(\omega-\frac{4qQ}{r}\right)^{2}+m_{0}^{2}f}}{f},
%\label{eqn:}
\end{equation}
where $+/-$ corresponds to the cases of the ingoing/outgoing fermion. In the work of Gwak \cite{sca-Gwak:2017kkt}, the positive sign in the above equation was selected. Thus, we get
\begin{equation}
p^{r}=g^{rr}p_{r}=\sqrt{\left(\omega-\frac{4qQ}{r}\right)^{2}+m_{0}^{2}f}.
%\label{eqn:}
\end{equation}
Near the event horizon, $f\rightarrow0$. The above equation is modified as
\begin{equation}
p^{r}=\omega-q\Phi,
\label{eqn:pr2}
\end{equation}
Therefore, the relation $\left(\ref{eqn:pr1}\right)$ can be recovered by fermion absorption. The above discussion mainly calculate the spin-up state. When it is the spin-down state, the result can be gotten is same as the spin-up state. It is clear that the results obtained by scalar particle absorption and fermion absorption are the same.
\section{Thermodynamics and particles¡¯ absorption with contribution of pressure and volume}
\label{sec:T}
In the extended phase space, the cosmological constant is treated as the function of the pressure of the black hole. After the black hole absorbs a particle, the changes of the internal energy and charge of the black hole equal to the energy and charge of the particle, that is
\begin{equation}
\omega=dU=d(M-PV), q=dQ.
\label{eqn:T dUdQ}
\end{equation}
Therefore, Eq. $\left(\ref{eqn:pr2}\right)$ is written as
\begin{equation}
dU=\frac{Q}{r_{+}}dQ+p^{r}.
\label{eqn:T pr}
\end{equation}
The initial state of the black hole is represented by $(M,Q,P,r_+)$, and the final state is
represented by $(M + dM, Q+dQ, P+dP, r_+ + dr_+)$. The function $f(M,Q,P,a,\alpha,r_{+})$ and $f(M+dM,Q+dQ,P+dP,a+da,\alpha+d\alpha,r_{+}+dr_{+})$ satisfy
\begin{equation}
f(M,Q,P,a,\alpha,r_{+})=f\left(M+dM,Q+dQ,P+dP,a+da,\alpha+d\alpha,r_{+}+dr_{+}\right)=0.
\label{eqn:T fE0}
\end{equation}
The relation between the functions $f(M,Q,P,a,\alpha,r_{+})$ and $f(M+dM,Q+dQ,P+dP,a+da,\alpha+d\alpha,r_{+}+dr_{+})$ is
\begin{equation}
\begin{aligned} & f\left(M+dM,Q+dQ,P+dP,a+da,\alpha+d\alpha,r_{+}+dr_{+}\right)=f(r)\\
 & +\frac{\partial f}{\partial M}|_{r=r_{+}}dM+\frac{\partial f}{\partial Q}|_{r=r_{+}}dQ+\frac{\partial f}{\partial r}|_{r=r_{+}}dr_{+}+\frac{\partial f}{\partial P}|_{r=r_{+}}dP+\frac{\partial f}{\partial a}|_{r=r_{+}}da+\frac{\partial f}{\partial\alpha}|_{r=r_{+}}d\text{\ensuremath{\alpha}},
\end{aligned}
\label{eqn:T f1eqn}
\end{equation}
where
\begin{equation}
\begin{aligned} & \frac{\partial f}{\partial M}|_{r=r_{+}}=-\frac{2}{r_{+}},\frac{\partial f}{\partial Q}|_{r=r_{+}}=\frac{2Q}{r_{+}^{2}},\frac{\partial f}{\partial r}|_{r=r_{+}}=4\text{\ensuremath{\pi}}T,\\
 & \frac{\partial f}{\partial P}|_{r=r_{+}}=\frac{8\pi r_{+}^{2}}{3},\frac{\partial f}{\partial\alpha}|_{r=r_{+}}=-\frac{1}{r_{+}^{3\omega_{q}+1}},\frac{\partial f}{\partial a}|_{r=r_{+}}=-1.\\
\end{aligned}
\label{eqn:ext T df}
\end{equation}
Using Eqs. $\left(\ref{eqn:T fE0}\right)$, $\left(\ref{eqn:T f1eqn}\right)$ and $\left(\ref{eqn:ext T df}\right)$ yields
\begin{equation}
dr_{+}=\frac{2p^{r}+r_{+}da+r_{+}^{-3\omega_{q}}d\alpha}{4\pi r_{+}(T-2Pr_{+})}.
\label{eqn:T dr}
\end{equation}
Then the variation of the entropy and volume is written as
\begin{equation}
dS=\frac{2p^{r}+r_{+}da+r_{+}^{-3\omega_{q}}d\text{\ensuremath{\alpha}}}{2T-4Pr_{+}},
\label{eqn:T dS}
\end{equation}
and
\begin{equation}
dV=\frac{2r_{+}p^{r}+r_{+}^{2}da+r_{+}^{-3\omega_{q}+1}d\alpha}{T-2Pr_{+}}.
\label{eqn:T dV}
\end{equation}
From Eqs. $\left(\ref{eqn:T}\right)$, $\left(\ref{eqn:T dS}\right)$, $\left(\ref{eqn:P}\right)$ and $\left(\ref{eqn:T dV}\right)$, we obtain
\begin{equation}
TdS-PdV=p^{r}+\frac{r_{+}}{2}da+\frac{r_{+}^{-3\omega_{q}}}{2}d\alpha.
\label{eqn:T TdSPdV}
\end{equation}
Moreover, substituting Eqs. $\left(\ref{eqn:potential}\right)$, $\left(\ref{eqn:T dUdQ}\right)$ and $\left(\ref{eqn:T pr}\right)$ into Eq. $\left(\ref{eqn:T TdSPdV}\right)$, the relation between the internal energy and enthalpy is simplified to
\begin{equation}
dM=TdS+VdP+\varPhi dQ+\mathcal{\gamma}d\alpha+\varkappa da,
\label{eqn:T dM r}
\end{equation}
where $\gamma$ and $\varkappa$ are the physical quantity conjugated
to the parameter $\text{\ensuremath{\alpha}}$ and $a$, respectively. They satisfy
\begin{equation}
\gamma=-\frac{1}{2r_{+}^{3\omega_{q}}},\varkappa=-\frac{r_{+}}{2}.
\end{equation}
Hence, the first law of thermodynamics is still satisfied.

When it is the extremal black hole, the temperature is zero. Then Eq. $\left(\ref{eqn:T dS}\right)$ is modified as
\begin{equation}
dS=-\frac{2p^{r}+r_{+}da+r_{+}^{-3\omega_{q}}d\text{\ensuremath{\alpha}}}{4Pr_{+}}.
\end{equation}
If $d\alpha>0$ and $da>0$, $dS$ is less than zero and the entropy of the extremal black hole decreases over time. If $d\alpha<0$ and $da<0$, $dS$ could be greater than zero. Therefore, the second law of thermodynamics is indefinite for the extremal black hole in the extended phase space.

When it is the near-extremal black hole, we will do numerically research on the change of entropy to intuitively understand the
changes in entropy. We set $M=0.5$ and $l=p^{r}=1$. For the case $\omega_{q}=-2/3$, $a=0.01$ and $\alpha=0.01$, the extremal charge is $Q_e=0.465706962$. When the charge is less than the extremal charge, different charge values are used to produce changes in entropy. The values of $r_+$ and $dS$ corresponding to different charge values are sorted out in Table \ref{tab:dS1}.
\begin{table}[h]
\centering
\begin{tabular}{ccccc}
\hline
\hline
$Q $               &$r_+ $               & $dS $       &$da$      &$d\alpha$  \\
\hline
0.465706962    & 0.388699            & $-12.1128$      &          &              \\
0.465706          & 0.389388           & $-12.2765$      &          &              \\
0.46              & 0.440516           & $-61.7333$     &          &              \\
0.44              & 0.495002          & $35.3525$      &          &              \\
0.42              & 0.495002           & $20.7666$      &0.5       &0.1        \\
0.40              & 0.550912            & $17.2452$      &          &              \\
0.30              & 0.623442          & $12.1402$      &          &             \\
0.20              & 0.661300           & $10.9798$      &          &              \\
0.10              & 0.680975           & $10.5455$       &          &              \\
\hline
\hline
\end{tabular}
\caption{The relation between $dS$, $Q$ and $r_+$.}
\label{tab:dS1}
\end{table}

From Table \ref{tab:dS1}, it can be obtained that the event horizon of the black hole and the variation of entropy increases when the charge of the black hole decreases. It is obviously that there are two regions where $dS>0$ and $dS<0$. Therefore, there exists a phase transition point that divides the value of $dS$ into positive and negative values regions.

In order to explore whether the values of state parameters of cloud of strings and quintessence have effects on the second law of thermodynamics, we test the validity of the second law when the values of $a$, $\alpha$ and $\omega_{q}$ change. In Table \ref{tab:dS2}, we set $a=0.1$ and other values of variables are the same as in Table \ref{tab:dS1}. The extremal charge is $Q_e=0.480782137$. Same as before, we use different charge values to produce changes in entropy. The values of $r_+$ and $dS$ corresponding to different charge values are summarized in Table \ref{tab:dS2}.
\begin{table}[htb]
\begin{centering}
\begin{tabular}{ccccc}
\hline
\hline
$Q $               &$r_+ $               & $dS $       &$da$      &$d\alpha$  \\
\hline
0.480782137    & 0.407715            & $-11.5960$      &          &              \\
0.480782          & 0.407979           & $-11.6471$      &          &              \\
0.48              & 0.427511           & $-16.6251$     &          &              \\
0.47              & 0.479200          & $-109.927$      &          &              \\
0.45              & 0.524836           & $45.8092$      &0.5       &0.1        \\
0.40              & 0.587213            & $19.7015$      &          &              \\
0.30              & 0.653150          & $14.1111$      &          &             \\
0.20              & 0.688780           & $12.6902$      &          &              \\
0.10              & 0.707491           & $12.1425$       &          &              \\
\hline
\hline
\end{tabular}
\par\end{centering}
\caption{{\footnotesize{}{}{}{}The relation between $dS$, $Q$ and $r_+$.}}
\label{tab:dS2}
\end{table}

In Table \ref{tab:dS3}, we set $\omega_{q}=-1/2$, $a=0.1$, $\alpha=0.1$ and other values of variables are the same as in Table \ref{tab:dS1}. The extremal charge is $Q_e=0.491516500$. The values of $r_+$ and $dS$ corresponding to different charge values are summarized in Table \ref{tab:dS3}.
\begin{table}[htb]
\begin{centering}
\begin{tabular}{ccccc}
\hline
\hline
$Q $               &$r_+ $               & $dS $       &$da$      &$d\alpha$  \\
\hline
0.491516500    & 0.424248           & $-12.5738$      &          &              \\
0.491516          & 0.424762           & $-12.6769$      &          &              \\
0.49              & 0.452203          & $-20.8312$     &          &              \\
0.47              & 0.525092         & $121.255$      &          &              \\
0.45              & 0.560603           & $37.4749$      &0.5       &0.1        \\
0.40              & 0.616308            & $20.7626$      &          &              \\
0.30              & 0.678621          & $15.3953$      &          &             \\
0.20              & 0.712972           & $13.8958$      &          &              \\
0.10              & 0.731128           & $13.3019$       &          &              \\
\hline
\hline
\end{tabular}
\par\end{centering}
\caption{{\footnotesize{}{}{}{}The relation between $dS$, $Q$ and $r_+$.}}
\label{tab:dS3}
\end{table}
\begin{figure}[htb]
\centering
\includegraphics[scale=0.8]{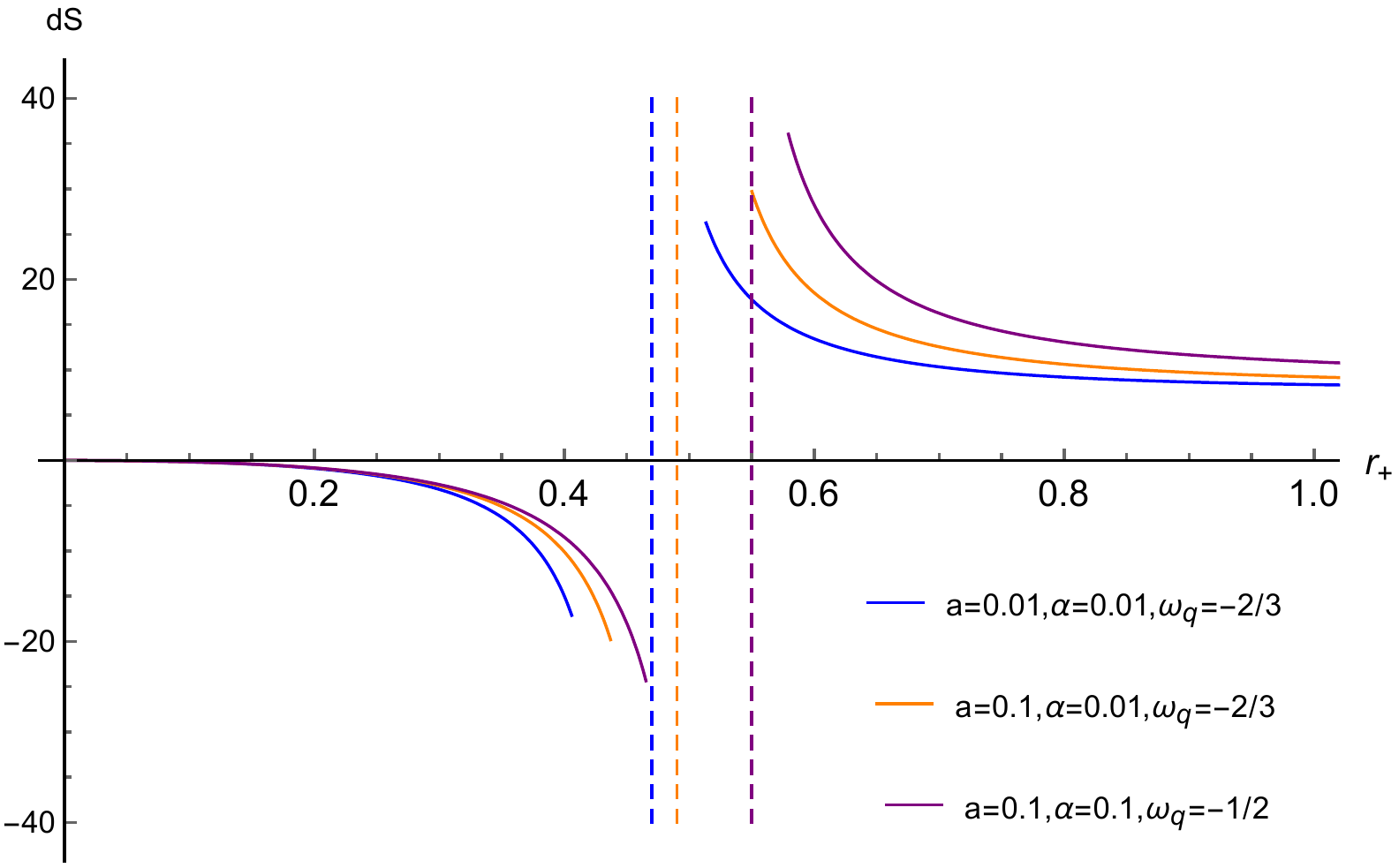}
\caption{The relation between $dS$ and $r_+$ which parameter values are $M=0.5$, $l=p^{r}=1$, $da=0.5$, and $d\alpha=0.1$.}
\label{fig:dS}
\end{figure}

In order to more intuitively observe the impact of $a$ and $\alpha$ on $dS$, the function graph is used to express the relationship between $dS$ and $r_+$ in different situations, which is shown in Fig. \ref{fig:dS}. From Fig. \ref{fig:dS}, it is clear that there is indeed a phase change point that divides $dS$ into positive and negative values. If the charge of the black hole is less than the extreme value of the charge, the change in entropy is negative and entropy decreases. If the charge is greater than the extreme charge, the change in entropy is positive and the entropy increases. Thus, the second law of thermodynamics is indefinite for the black hole in the extended phase space. From Table \ref{tab:dS1}, \ref{tab:dS2} and \ref{tab:dS3}, we find when the value of state parameter of cloud of strings or quintessence increase, extremal charge $Q_e$ and its corresponding $r_+$ also increase. Moreover, for the same value of $r_+$, $dS$ increases when $a$, $\alpha$ and $\omega_q$ increase. The values of the parameters do have effects on the second law of thermodynamics, but the parameters do not determine whether the second law of thermodynamics is ultimately violated.

\begin{figure}[h]
\centering
\includegraphics[scale=0.65]{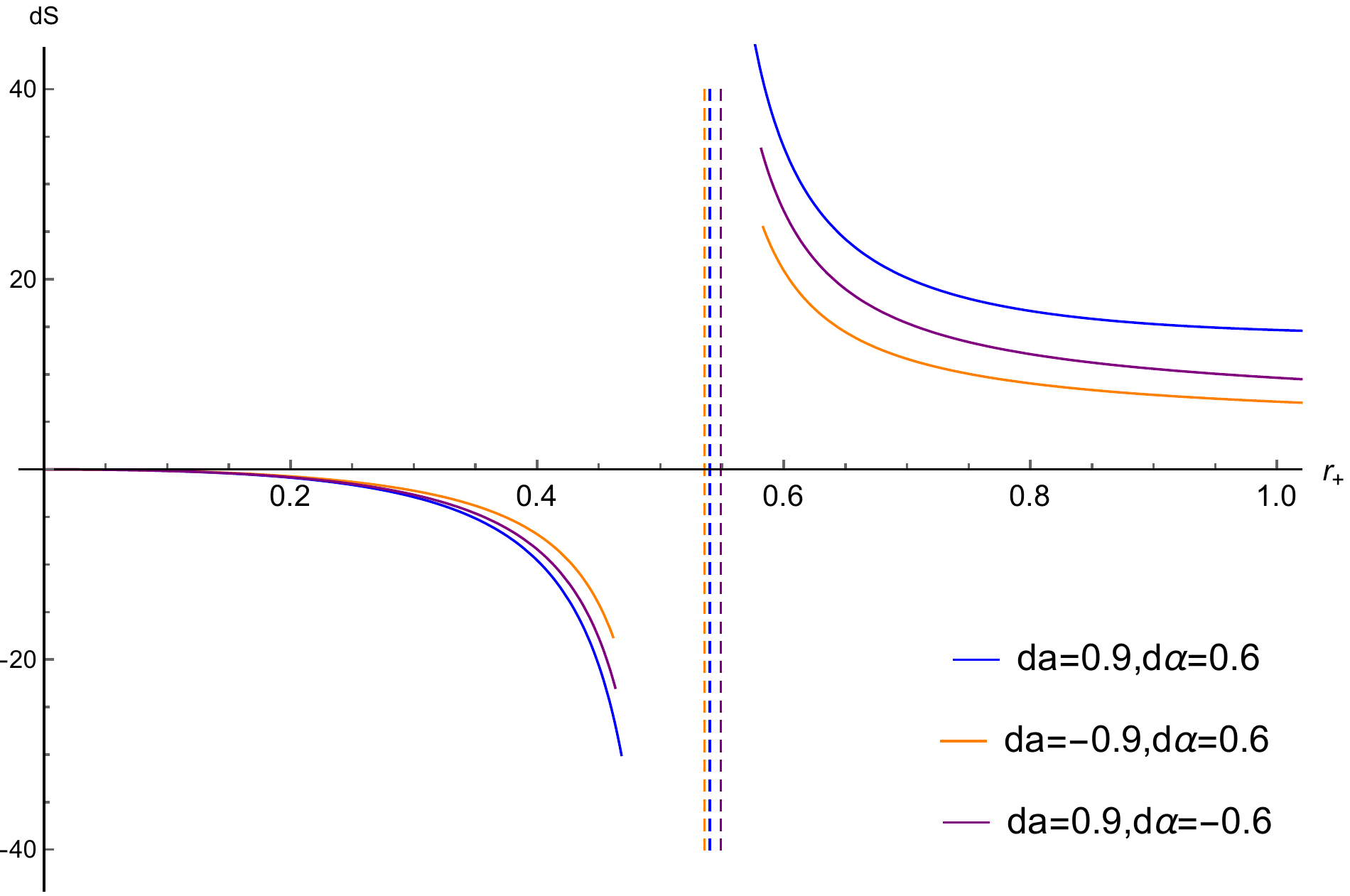}
\caption{The relation between $dS$ and $r_h$ which parameter values are $M=0.5$, $l=p^{r}=1$, $\omega_q=-2/3$, $a=0.1$, and $\alpha=0.1$.}
\label{fig:dS1}
\end{figure}
Furthermore, $da$ and $d\alpha$ also have effects on $dS$. When $da$ and $d\alpha$ change, we numerically analyze the change of $dS$ and plot it, as shown in Fig. \ref{fig:dS1}. From the above discussion, it is concluded that the second law of thermodynamics is not always valid for near-extremal black holes in the extended phase space.

\section{Overcharging problem in the extremal and near-extremal black holes}
\label{sec:wccc}
The validity of WCCC in the extremal and near-extremal black holes is discussed by the absorptions of the scalar particle and fermion in this section. An effective way to test the validity of WCCC is to check whether the event horizon exists after the black hole absorbs the particles. The event horizon is determined by the function $f(r)$. In the initial state, the minimum value of $f(r)$ is negative or zero and $f(r)=0$ has real roots. It means event horizon exists. When the black hole absorbs a particle, the mass and charge of the black hole will change during the infinitesimal time interval $dt$ and the minimum value of $f(r)$ will also change. If the minimum value of $f(r)$ is negative or equal to zero, the event horizon exists. Therefore the WCCC is effective. Otherwise, the minimum value of $f(r)$ is positive, the event horizon doesn't exist. In this case, the black hole is overcharged and the WCCC is ineffective.

The sign of the minimum value in the final state can be obtained in term of the initial state \cite{WCCC-Gwak:2019asi}. Assuming $(M,Q,P,r_0,a,\alpha)$ and $(M+dM,Q+dQ,P+dP,r_0+dr_0,a+da,\alpha+d\alpha)$ represent the initial state
and the finial state, respectively. At $r=r_0+dr_0$, $f(M+dM,Q+dQ,P+dP,r_0+dr_0,a+da,\alpha+d\alpha)$ is written as
\begin{equation}
\begin{aligned}
&f\left(M+dM,Q+dQ,P+dP,a+da,\alpha+d\text{\ensuremath{\alpha}},dr_{0}+r_{0}\right)\\
&=\delta+\frac{\partial f}{\partial M}|_{r=r_{0}}dM+\frac{\partial f}{\partial Q}|_{r=r_{0}}dQ+\frac{\partial f}{\partial P}|_{r=r_{0}}dP\\
&+\frac{\partial f}{\partial a}|_{r=r_{0}}da+\frac{\partial f}{\partial\alpha}|_{r=r_{0}}d\text{\ensuremath{\alpha}}+\frac{\partial f}{\partial r}|_{r=r_{0}}dr,\\
\end{aligned}
\label{eqn:W f1}
\end{equation}
where
\begin{equation}
\begin{aligned}
&\frac{\partial f}{\partial r}|_{r=r_{0}}=0,\frac{\partial f}{\partial M}|_{r=r_{0}}=-\frac{2}{r_{0}},\frac{\partial f}{\partial Q}|_{r=r_{0}}=\frac{2Q}{r_{0}^{2}},\\
&\frac{\partial f}{\partial P}|_{r=r_{0}}=\frac{8\text{\ensuremath{\pi}}r_{0}^{2}}{3},\frac{\partial f}{\partial a}|_{r=r_{0}}=-1,\frac{\partial f}{\partial\alpha}|_{r=r_{0}}=-\frac{1}{r_{0}^{3\omega_{q}+1}}.\\
\end{aligned}
\label{eqn:W df}
\end{equation}
At $r=r_0$, we have
\begin{equation}
f\left(M,Q,P,a,\alpha,r_{0}\right)\equiv f_{0}=\delta\leq0,
\label{eqn:W 1}
\end{equation}
and
\begin{equation}
\partial_{r}f\left(M,Q,P,a,\alpha,r_{0}\right)\equiv f_{min}^{\prime}=0.
%\label{eqn:W f1}
\end{equation}
From Eqs. $\left(\ref{eqn:W f1}\right)$, $\left(\ref{eqn:W df}\right)$, $\left(\ref{eqn:T dUdQ}\right)$, $\left(\ref{eqn:T pr}\right)$ and $\left(\ref{eqn:T dV}\right)$, we obtain
\begin{equation}
\begin{aligned}
&f\left(M+dM,Q+dQ,P+dP,a+da,\alpha+d\text{\ensuremath{\alpha}},r_{0}+dr_{0}\right)\\
&=\delta-\frac{2Tp^{r}}{r_{0}\left(T-2Pr_{+}\right)}-\frac{2qQ}{r_{0}}\left(\frac{1}{r_{+}}-\frac{1}{r_{0}}\right)+\frac{8\pi}{3r_{0}}\left(r_{+}-r_{0}\right)dP\\
&+\frac{Tr_{0}^{-3\omega_{q}}+2Pr_{+}\left(r_{+}^{-3\omega_{q}}-r_{0}^{-3\omega_{q}}\right)}{r_{0}\left(T-2Pr_{+}\right)}d\ensuremath{\alpha}+\frac{Tr_{0}+2Pr_{+}\text{(}r_{+}-r_{0})}{r_{0}\left(T-2Pr_{+}\right)}da.\\
\end{aligned}
\label{eqn:W f2}
\end{equation}

When the initial black hole is the extremal black hole, $r_{0}=r_{+}$, $T=0$ and $\delta=0$. Then we can obtain
$f_{min}=\delta=0$ and $f_{min}^{\prime}=0$. Hence, Eq. $\left(\ref{eqn:W f2}\right)$
is written as
\begin{equation}
f\left(M+dM,Q+dQ,P+dP,a+da,\alpha+d\text{\ensuremath{\alpha}},dr_{0}+r_{0}\right)=0.
%\label{eqn:W f1}
\end{equation}

When the initial black hole is the near-extremal black hole, $r_0$ and $r_+$ do not coincide. In Eq. $\left(\ref{eqn:W f2}\right)$, the first term in the second line satisfies Eq. $\left(\ref{eqn:W 1}\right)$ and the second term is only suppressed by the test particle limit. However, the other terms are suppressed by the approaching extreme value limit and the test particle limit. Therefore, they can be ignored. Hence, Eq. $\left(\ref{eqn:W f2}\right)$ is modified as
\begin{equation}
f\left(M+dM,Q+dQ,P+dP,a+da,\alpha+d\text{\ensuremath{\alpha}},r_{0}+dr_{0}\right)=\delta-\frac{2Tp^{r}}{r_{0}\left(T-2Pr_{+}\right)}<0.
%\label{eqn:W f1}
\end{equation}
Hence, the near-extremal black hole stays near-extremal after absorbing a charged particle. The WCCC is satisfied for both the extremal and near-extremal black holes in the extended phase space.

\section{Conclusion}
\label{sec:con}
In this paper, we investigated the first and second laws of the thermodynamics and the overcharging problem in a RN-AdS black hole with cloud of strings and quintessence via the absorptions of the scalar particle and fermion in the extended phase space. The cosmological constant is treated as the function of thermodynamic pressure $P$. Moreover, the state parameters of cloud of strings and quintessence are treated as variables. To study the variations of the thermodynamic quantities of the black hole after absorbing a charged particle, we calculated the absorption of scalar particle and fermion. We found they finally simplified to the same relation $p^{r}=\omega-q\Phi$. This relation is exactly the same as that obtained by the Hamilton-Jacobi equation. The reason is that the Hamilton-Jacobi equation can be obtained by inserting the wave function $\left(\ref{eqn:Psi}\right)$ into the Klein-Gordon equation $\left(\ref{eqn:KG}\right)$. The Hamilton-Jacobi equation takes on the form as
\begin{equation}
\left(\partial^{\mu}I+eA^{\mu}\right)\left(\partial_{\mu}I+eA_{\mu}\right)-m^{2}=0.
%\label{eqn:pr1}
\end{equation}
In addition, Hamilton-Jacobi equation can be obtained from Dirac equation \cite{con-Benrong:2014woa,con-Erbin:2017zwo}. The multiplication of Eq. $\left(\ref{eqn:Dirac}\right)$ gives the second-order partial derivative equation. The wave function is written as $\varPsi_{F}=\varPsi_{0}e^{\frac{i}{h}I}$, where $\varPsi_{0}$ is a position-dependent spinor. By inserting the function into the second-order partial derivative equation and keeping only the first-order term in $\hbar$, the Hamilton-Jacobian equation can be obtained.

We used the relation $p^{r}=\omega-q\Phi$ to recovered the first law of thermodynamics and discussed the validity of the second law of thermodynamics and WCCC. During the discussion, the final state of the black hole was considered to be still a black hole. The first law of thermodynamics is recovered and the second law of thermodynamics is indefinite. Since we treated the parameters related to cloud of strings and quintessence as variables, there are two more terms, $-\frac{1}{2r_{+}^{3\omega_{q}}} d\alpha$ and $-\frac{r_{+}}{2} da$, in the formula of the first law of thermodynamics. The detailed formulas of the first law of thermodynamics for black holes and black holes surrounded by clouds and quintessence are shown in Table \ref{tab:1st}.
\begin{table}[htb]
\begin{centering}
\begin{tabular}{|p{3.0in}|p{3.0in}|}
\hline
Types of black holes & 1st law in the extend phase space  \tabularnewline
\hline
RN-AdS BH  & $dM=TdS+VdP+\text{\ensuremath{\varphi}}dQ$\tabularnewline
\hline
RN-AdS BH with cloud of strings and quintessence  & $dM=TdS+VdP+\text{\ensuremath{\varphi}}dQ-\frac{1}{2r_{+}^{3}\omega_{q}}d\alpha-\frac{r_{+}}{2}da$\tabularnewline
\hline
\end{tabular}
\par\end{centering}
\caption{{\footnotesize{}{}{}{}Results for the first
thermodynamic law under different conditions in the extend phase space.}}
\label{tab:1st}
\end{table}

When tested the validity of the second law of thermodynamics, we found there exists a phase change point that divides the value of $dS$ into positive and negative regions. The change in the values of the state parameters related to the cloud of strings and quintessence will affect the value of $r_+$, $Q_e$ and $dS$. As shown in Fig. \ref{fig:dS}, the parameters do have effects on the second law of thermodynamics. But the parameters do not determine whether the second law of thermodynamics is ultimately violated. Moreover, the value of $da$ and $d\alpha$ also affect the value of $Q_e$ and $dS$. Furthermore, the WCCC has been proven to be valid all the time for extremal and near-extremal black holes. The validity of WCCC was tested by checking the sign of the minimum value of $f(r)$. Compared with black holes without cloud of strings and/or quintessence, the minimum value of $f(r)$ becomes larger after absorbing particles with energy and charge. But the minimum value of $f(r)$ remains the original positive and negative. Therefore, neither extreme black holes or near-extreme black holes will be overcharged. Our results are shown in Table \ref{tab:12wccc}.
\begin{table}[htb]
\begin{centering}
\begin{tabular}{|p{0.9in}|p{3.7in}|}
\hline
1st law    & Satisfied\tabularnewline
\hline
2nd law    & Indefinite\tabularnewline
\hline
WCCC    & Satisfied for the extremal and near-extremal black holes. The extremal/near-extremal black hole stays extremal/near-extremal after the charge test particle absorption\tabularnewline
\hline
\end{tabular}
\par\end{centering}
\caption{{\footnotesize{}{}{}{}Results for the first and second laws of
thermodynamics and the WCCC, which are tested for a RN-AdS black hole with cloud of strings
and quintessence via the charge test particle absorption.}}
\label{tab:12wccc}
\end{table}

\begin{acknowledgments}
We are grateful to Peng Wang, Haitang Yang,Deyou Chen and Xiaobo Guo for useful discussions. This
work is supported in part by NSFC (Grant No. 11747171), Natural Science
Foundation of Chengdu University of TCM (Grants nos. ZRYY1729 and ZRYY1921),
Discipline Talent Promotion Program of /Xinglin Scholars(Grant no.
QNXZ2018050), the key fund project for Education Department of Sichuan (Grant
no. 18ZA0173).
\end{acknowledgments}

\end{document}